\documentclass[
 aip,
 apl,
 amsmath,amssymb,
 preprint,%
 ]{revtex4-2}

\usepackage{graphicx}
\usepackage{dcolumn}
\usepackage{bm}
\usepackage{siunitx}
\usepackage{xcolor}
\usepackage[utf8]{inputenc}
\usepackage[T1]{fontenc}
\usepackage{mathptmx}
\usepackage{amsmath}
\usepackage{etoolbox}

\usepackage{empheq, cases}

\makeatletter
\def\@email#1#2{%
 \endgroup
 \patchcmd{\titleblock@produce}
  {\frontmatter@RRAPformat}
  {\frontmatter@RRAPformat{\produce@RRAP{*#1\href{mailto:#2}{#2}}}\frontmatter@RRAPformat}
  {}{}
}%
\makeatother

\makeatletter
\let\@fnsymbol\@fnsymbol@latex
\@booleanfalse\altaffilletter@sw
\makeatother

\graphicspath{{./Figures/}}

\newcommand{\highlight}[1]{\textcolor{black}{#1}}
\newcommand{\edit}[1]{\textcolor{blue}{#1}}

\begin{document}


\title[Thermoelectric Fingerprinting of Bloch- and N\'{e}el-type Skyrmions]{Thermoelectric Fingerprinting of Bloch- and N\'{e}el-type Skyrmions}
	\author{Christopher E. A. Barker(*)}
	\email[]{christopher.barker@npl.co.uk}
	\affiliation{National Physical Laboratory, Hampton Road, Teddington, TW11 0LW, United Kingdom}
	
	\author{Elias Saugar}
	\affiliation{Instituto de Ciencia de Materiales de Madrid, ICMM–CSIC, Campus de Cantoblanco, C. Sor Juana In\'{e}s de la Cruz, 3, Madrid 28049, Spain}
	
	\author{Katharina Zeissler}
	\affiliation{National Physical Laboratory, Hampton Road, Teddington, TW11 0LW, United Kingdom}
	\affiliation{School of Physics and Astronomy, University of Leeds, Leeds LS2 9JT, United Kingdom}
	\affiliation{Bragg Centre for Materials Research, University of Leeds, Leeds LS2 9JT, United Kingdom}
	
	\author{Robert Puttock}
	\affiliation{National Physical Laboratory, Hampton Road, Teddington, TW11 0LW, United Kingdom}
	
	\author{Petr Klapetek}
	\affiliation{Czech Metrology Institute, Okruzni 772/31, Brno 10135, Czech Republic}
	\affiliation{Central European Institute of Technology (CEITEC), Brno University of Technology, Purkynova 123, Brno 612 00, Czech Republic.}

	\author{Olga Kazakova}
	\affiliation{National Physical Laboratory, Hampton Road, Teddington, TW11 0LW, United Kingdom}
	\affiliation{Department of Electrical and Electronic Engineering, University of Manchester, Manchester, M13 9PL, United Kingdom}
	
	\author{Christopher H. Marrows(*)}
	\affiliation{School of Physics and Astronomy, University of Leeds, Leeds LS2 9JT, United Kingdom}
	
	\author{Oksana Chubykalo-Fesenko}
	\affiliation{Instituto de Ciencia de Materiales de Madrid, ICMM–CSIC, Campus de Cantoblanco, C. Sor Juana In\'{e}s de la Cruz, 3, Madrid 28049, Spain}
	
	\author{Craig Barton(*)}
	\email[Author to whom correspondence should be addressed: ]{craig.barton@npl.co.uk, C.H.Marrows@leeds.ac.uk}
	\affiliation{National Physical Laboratory, Hampton Road, Teddington, TW11 0LW, United Kingdom}
	
	\date{\today}

\date{\today}

\begin{abstract}
Magnetic skyrmions are nanoscale spin textures that exhibit topological stability, which, along with their thermal and electrical transport properties, make them the ideal candidates for a variety of technological applications.
Accessing the skyrmion spin texture at the nanoscale and understanding its interaction with local thermal gradients is essential for engineering skyrmion-based transport phenomena. 
However, direct experimental insight into the local thermoelectric response of single skyrmions remains limited. To address this, we employ scanning thermoelectric microscopy~(SThEM) to probe the nanoscale thermoelectric response from a single skyrmion. 
By mapping the local thermoelectric voltage with nanoscale precision, we reveal a unique spatially resolved response that is the convolution of the underlying spin texture of the skyrmion and its interaction with the highly localised thermal gradient originating from the heated probe. 
We combine this with thermoelectric modelling of a range of skyrmion spin textures to reveal unique thermoelectric responses and allow the possibility of SThEM to be used as a tool to distinguish nanoscale spin textures. 
These findings provide fundamental insights into the interaction of topologically protected spin textures with local thermal gradients and the resultant spin transport. 
We demonstrate a route to thermally characterise nanoscale spin textures, accelerating the material optimisation cycle, while also opening the possibility to harness skyrmions for spin caloritronics.       
\end{abstract}

\maketitle

Magnetic skyrmions are topologically protected spin textures that can emerge in chiral magnetic systems due to the interplay between the Dzyaloshinskii--Moryia interaction (DMI), perpendicular magnetic anisotropy, exchange interactions and external magnetic fields~\cite{dzyaloshinskii1965theory, Bogdanov1989, 10.1063/1.4943757, nagaosa2013topological, wiesendanger2016nanoscale, fert:hal-01695591, 10.1063/1.5048972, Zhang_2020, gobel2021beyond}. 
Materials with these nanoscale spin textures~\cite{PhysRevLett.114.177203, 10.1038/s41467-019-11831-4, 10.1038/s41565-018-0255-3} exhibit unique electrical transport properties, such as the topological Hall effect~\cite{PhysRevB.92.115417, WANG2022100971}. 
Furthermore, owing to their strong coupling to spin currents and Rashba interfaces, current driven motion can occur at very low current densities~\cite{10.1063/1.4943757, Jiang2016, Legrand2017, Juge2019, Zeissler2020, Litzius2020}. 
These properties make them exciting candidates for a future generation of spintronic based logic and storage technologies~\cite{Wang2022}. 
Furthermore, they are also attractive candidates for non von Neumann computational paradigms such as neuromorphic computing~\cite{Lee2023}. 
The local spin texture plays a fundamental role in governing skyrmion dynamics; interactions; and their response to external stimuli such as electric fields~\cite{Li2022}, currents~\cite{Buttner2017, WANG2022100971} and thermal gradients~\cite{Yu2021, Gong2022}. 
Understanding the nanoscale spin structure is crucial for controlling skyrmion-based transport phenomena, including the interplay between spin-orbit effects~\cite{Jiang2016}, their thermoelectric response~\cite{Scarioni2021} and magnon-mediated transport~\cite{Li2022JAP}. 
Hence, the ability to understand and measure the local spin texture is of great importance, both from a fundamental perspective, and also to optimise device performance during material development; creating a route to expedite the \emph{lab--to--fab} development of next generation skyrmion devices.\\
Probing the local properties of nanoscale spin textures requires high-resolution microscopy techniques capable of detecting their local electronic, magnetic, or---as in our work---thermoelectric response. 
There have been significant developments in the real-space characterisation of nanoscale spin textures, where a range of measurement solutions are available to ascertain the local magnetisation with remarkable precision and sensitivity. 
These techniques include: x-ray magnetic nanotomography; Lorentz transmission electron microscopy (LTEM); spin-polarized scanning tunneling microscopy (SPSTM); scanning probe nitrogen--vacancy~(NV) magnetometry; and magnetic force microscopy~(MFM)~\cite{Romming2015, PhysRevLett.114.177203, McVittie2018, Dovzhenko2018, Legrand2018, Gross2018, Finco2021, Donnelly2018, Donnelly2020, Hug1998, Schendel2000, Bacani2019, Yagil2018, Barton2023}. 
More recently, scanning thermoelectric microscopy~(SThEM) has emerged as powerful technique to map the the various local thermoelectric responses with nanometre resolution, enabling direct investigation of the coupling between spin textures and thermal gradients~\cite{Bartell2015, Pfitzner2018, Iguchi2019, Gray2019, Janda2020, Sola2020, Harzheim2020, Puttock2022} providing an invaluable tool to study spin caloritronic effects at the nanoscale~\cite{Bauer2012, Boona2014, Yu2017}. 
In this work, we employ SThEM to measure the thermoelectric signature of an individual skyrmion, providing insight into the fundamental interactions that govern skyrmionic thermoelectric effects at the nanoscale. 
A combination of experimental and thermoelectric modelling is used to distinguish the signal resulting from two skyrmion types: Bloch and N\'{e}el. 
Our findings provide the foundations to ascertain the local real space magnetisation of topological spin textures using a generic lab--based scanning probe microscope, and, more fundamentally, offer a route to explore thermally driven charge, spin, and thermal~(magnon) transport in complex magnetic spin textures at the nanoscale.\\\\
\begin{figure}[!ht]
\includegraphics[width=0.42\textwidth]{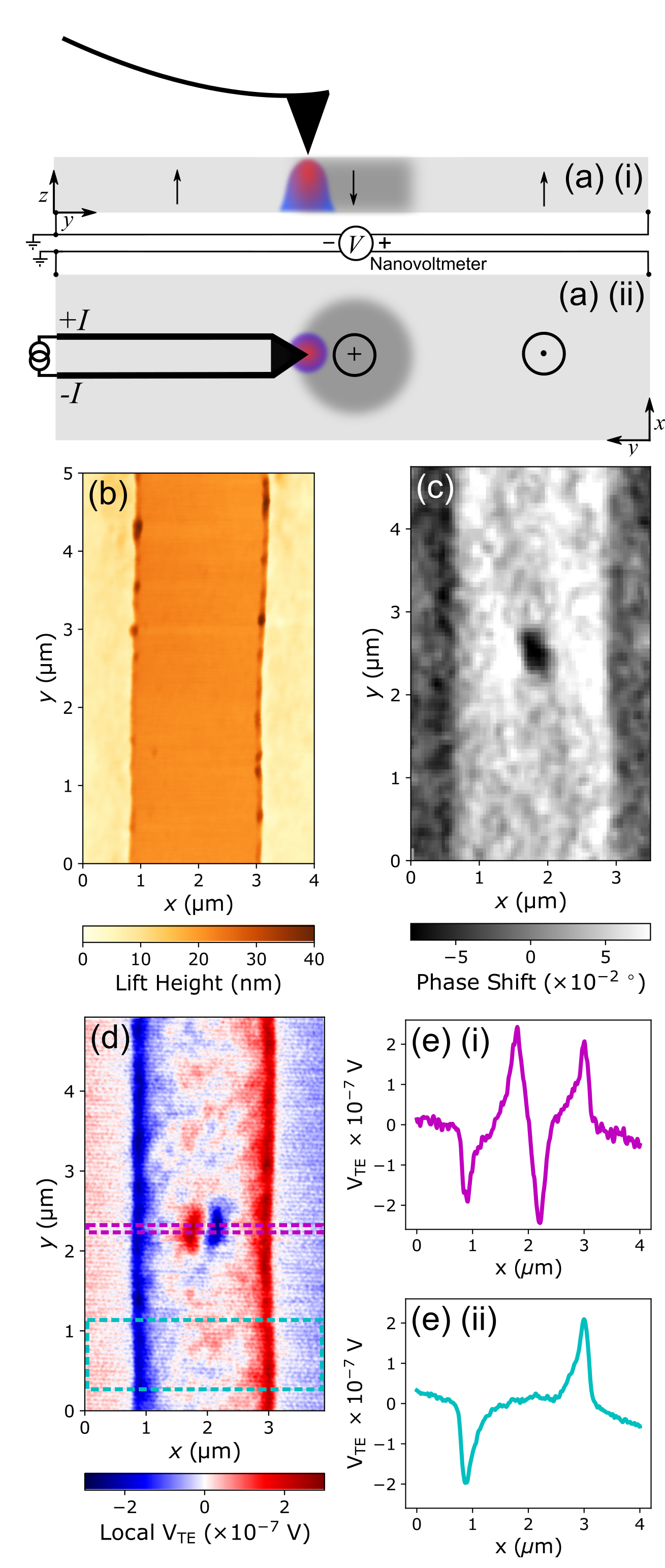}
\caption{(a) Conceptual overview of scanning thermoelectric microscopy, showing schematic representation of the top (i) and side (ii) views, respectively. 
	Dark grey regions represent the the position of the skyrmion and highlight the local projection of the z--component of magnetisation. 
	Coordinate system used for the modelling is also shown. 
	(b) Atomic force micrograph showing the topography of the microwire. 
	(c) Magnetic force micrograph showing a single skyrmion nucleated in the microwire. 
	(d) Scanning thermoelectric micrograph showing the thermoelectric response from a single skyrmion, and (e) (i,ii) averaged line profiles taken from the data in (d) from skyrmion centre (i) and saturated (ii) region, respectively.}
\label{fig:main_one}
\end{figure}
Fig.~\ref{fig:main_one} (a) schematically shows the SThEM experimental setup, detailing how a constant--current Joule heated~(ThermaLever\texttrademark{}) scanning probe, \highlight{operated in contact mode}, is used to introduce a nanometric thermal gradient. 
Electrical readout was achieved using a Keysight 34420A nanovoltmeter which was synced to the scanning probe microscopes controller to such that the electrical readout could be mapped spatially with the sample topography~\cite{Puttock2022}. \highlight{The device contact pads~(outside of the field of view presented here and separated by $\approx$40 $\mu$m) were connected to the nanovoltmeter via a PCB and breakout box and the input low was connected to the microscopes ground.}
Fig. 1 (a) (i,ii) shows representations of the top and side views of the sample. 
In order to stabilise skyrmions in our system, a sizable interfacial DMI is required. 
To meet these requirements, a perpendicular magnetic anisotropy~(PMA) asymmetric multilayer comprising three repeats of a trilayer, including buffer and cap, where the composition was Ta(3.2)/Pt(2.7)[Co$_{68}$B$_{32}$(0.8)/Ir(0.4)/Pt(0.6)]$_{x3}$/Pt(2.2) (thicknesses in nm). 
The multilayer structure was sputter deposited using an ultra high vacuum deposition system, analogous to previously studied samples\highlight{, where it was shown that N\'eel-type skyrmions are stabilised.}~\cite{Finizio2019, Zeissler2020, Barker2025}. 
Patterning of the multilayer was done using standard optolithogrpahy and lift-off to create a 2.3~$\mu$m wide microwire including the voltage contact pads required for electrical readout of thermoelectric signals. 
Fig.~\ref{fig:main_one} (b) shows an atomic force micrograph detailing the topography of the microwire used in this work. 
Fig.~\ref{fig:main_one} (c) shows a magnetic force micrograph and confirms the presence of singular skyrmion spin texture of $\lessapprox$~200 nm in radius. 
Skyrmion nucleation was achieved using the stray magnetic field from the MFM tip~\cite{Barker2025}. 
The corresponding zero-field scanning thermoelectric micrograph is shown in Fig.~\ref{fig:main_one} (d), along with line profiles in Fig.~\ref{fig:main_one}(e)(i,ii).

\highlight{The data shown here has been treated with a simple FFT filter to remove high-frequency periodic noise, along with a 2 pixel Gaussian filter to improve the data presentation. Furthermore, the offset present in the data was removed to zero the background signal, where no thermoelectric response is expected to arise.} 

Here, we observe several characteristic features in the local thermoelectric response: the response from the device edges~\cite{Puttock2022}, shown in Fig.~\ref{fig:main_one}(e)(i), and the lobe-like features that result from the skyrmion itself, Fig.~\ref{fig:main_one}(e)(ii), both of which are discussed in detail below. 
We note that the zero field stability is likely to occur due to the attended pinning in these multilayer material systems.\\
To understand the local thermoelectric response in more detail and we have modelled the response for  Bloch and N\'{e}el type skyrmions.
\highlight{Additional measurements of zero-field skyrmions in devices made of the same and similar multilayers are included in the supplementary material.}
The model approximates the magnetic system as a single magnetic layer, where the spin textures have been analytically defined and constitutes the input to the semi--analytical thermoelectric calculation. 
The thermal response to a heated tip was simulated using a custom finite difference solver~\cite{Klapetek2017}, where from previous work~\cite{Puttock2022}, the temperature at the tip--sample contact point was assumed to be $\approx$327~K. \highlight{An example of the simulated thermal gradient, and its corresponding profile, for a simple 2-dimensional system is presented in the supplementary material.} 
We note that it is the spatial extent of the temperature profile~(the effective point spread function of the technique), which when convolved with the magnetisation texture, defines the localisation of the resultant electrical field and the resolution.  
The electric field $\mathbf{E}$ that results from a thermal gradient $\nabla \mathbf{T}$ is defined by the Seebeck tensor $S$ through $\mathbf{E} = S\nabla \mathbf{T}$. 
$S$ is constructed to contain all sources of thermoelectric response that we anticipate in our system~\cite{Reimer2017, Krzysteczko2017}. 
It includes the anomalous Nernst effect~(ANE); the anisotropic magneto-thermopower~(AMTP); and the planar Nernst effect~(PNE).\\
These effects result from the asymmetry in the energy dependent scattering of electrons and holes at the Fermi level, where their differential density of states and velocities yields a net thermoelectric current~\cite{Bauer2012} which is spin polarised and interacts with the local spin texture. 
The ANE accounts for the transverse electric field that arises in ferromagnetic materials with spin--orbit coupling when subjected to a thermal gradient, which originates from a magnetisation dependent Berry curvature and extrinsic scattering~\cite{Smit1955, Berger1970, Berry1984,Nagaosa2010}. 
The AMTP and PNE account for the thermoelectric response due to spin-dependent scattering and band structure orbital anisotropy which is highly sensitive to the angle between the local magnetisation and thermal gradient~\cite{Campbell1970} and occurs when $S\nabla \mathbf{T}$ is coplanar with the local magnetisation. 
This anisotropy is captured by the difference between the in-plane perpendicular and parallel Seebeck coefficients, $S_{\perp}\ \&\ S_{\parallel}$, respectively. 
Including these effects in $S$, Eqn.~\ref{Eqn:-1}, allows the expression for the total electric field $\mathbf{E}$ for an arbitrary magnetisation direction to be derived. 
This is achieved by performing the appropriate rotation operations on $S$ allowing the analytical expressions of the thermoelectric response to be computed for different planes of rotation of magnetisation Eqn.~\ref{Eqn:1}, in addition to the measurement geometries. \highlight{For a uniform out--of--plane magnetisation $S$ is given by:}

\begin{equation}
S = \begin{pmatrix}
S_{\perp} & -S_N & 0  \\
S_N& S_{\perp} & 0 \\
0 &0  & S_{\parallel}
\end{pmatrix} ,
\label{Eqn:-1}
\end{equation}

In this work, we are confined to the electrical measurement along the length of the device $\hat{y}$, and, as stated, responses due to in-plane thermal gradients. 
The resulting electric field, for an arbitrary configuration is given by:    
\begin{equation}
 \mathbf{E} = S_{\perp} \nabla \mathbf{T} + (S_{\parallel} - S_{\perp})(\nabla \mathbf{T}\cdot \hat{\mathbf{m}})\hat{\mathbf{m}} -  S_{N} (\hat{\mathbf{m}}\times    \nabla \mathbf{T}),
\label{Eqn:1}
\end{equation}
where $S_{N}$ it the anomalous Nernst Seebeck coefficient, and $\nabla \mathbf{T}$ and $\hat{\mathbf{m}}$ are the thermal gradient and magnetisation unit vector respectively.\\
The voltage calculated along the length, between measurement points $A$ and $B$, and across the width of the device $w$, is simply computed using $V_{y} = -\int^{B}_{A} \mathbf{E}\cdot dl$. 
In this work we have included only thermoelectric responses that results from in plane thermal gradients and discounted those occurring from vertical temperature differences \highlight{due to the small thickness of our sample~\cite{Puttock2022, Uchida2008, Adachi_2013}}, i.e. 
$\nabla \mathbf{T} = (\nabla T_{x}, \nabla T_{y}, 0 )$. 
\begin{figure*}[t!]
\includegraphics[width=0.95\textwidth]{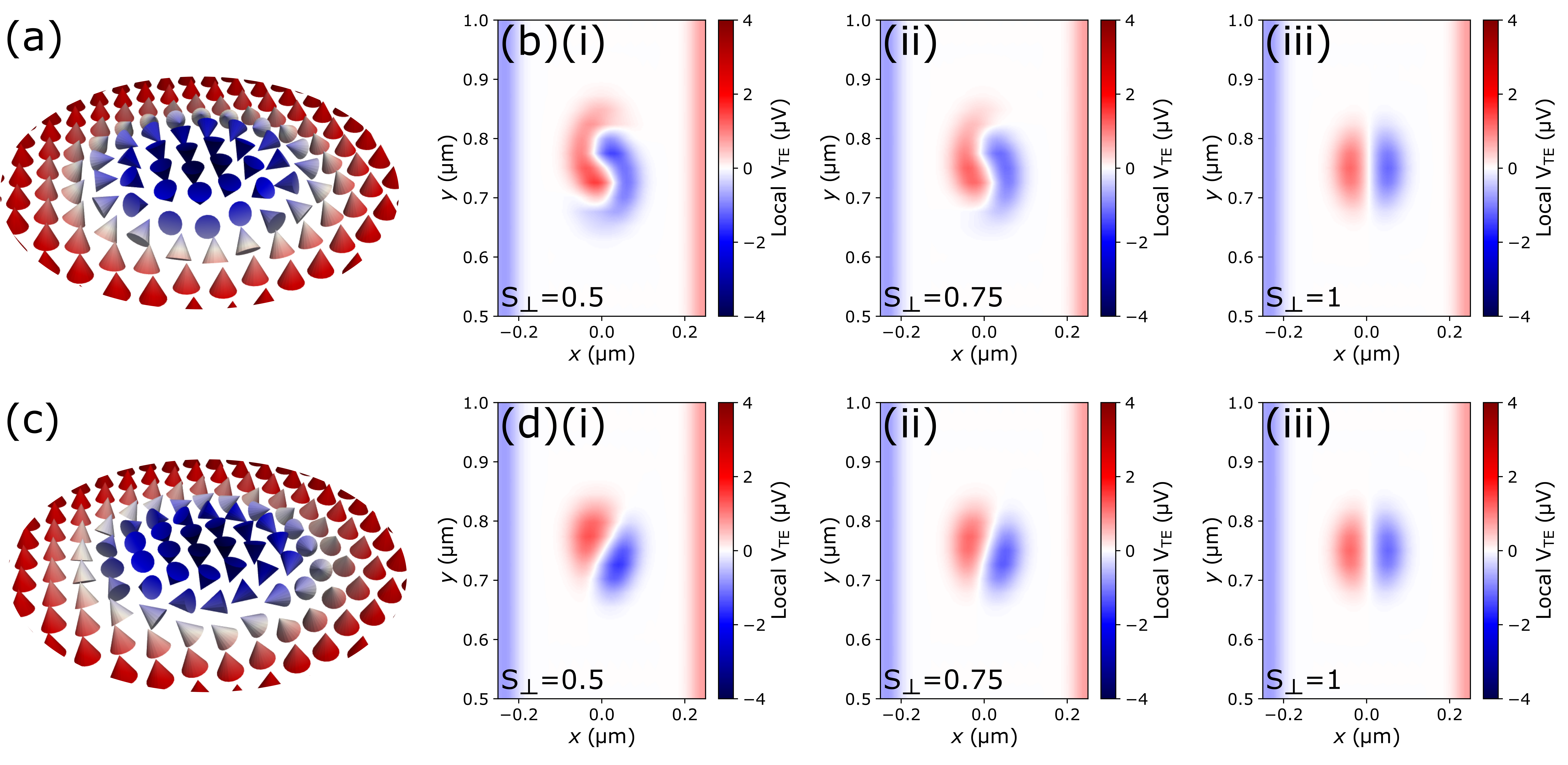}
\caption{Shows modelling of the thermoelectric response of a single skyrmion. 
	(a) shows the analytically defined N\'{e}el spin texture used as the input for the thermoelectric modelling. 
	(b)(i--iii) show the total thermoelectric response for the N\'{e}el skyrmion for $S_{ \parallel } = -1.0$ and $S_{ \perp} =$ -0.5 (i), -0.75 (ii), and -1.0 (iii)~$\mu\mathrm{V}\cdot \mathrm{K}^{-1}$. 
	(c,d) Show the same for a Bloch skyrmion. 
	(c) is the analytically defined spin texture, and (d)(i--iii) again correspond to $S_{ \perp} =$ -0.5 (i), -0.75 (ii), and -1.0 (iii)~$\mu\mathrm{V}\cdot \mathrm{K}^{-1}$.}
\label{fig:main_two}
\end{figure*}
It is important to understand the role of the balance of the in-plane Seebeck coefficients as these will couple to the in-plane spin components and yield the most valuable information for fingerprinting the Bloch or N\'{e}el character of the skyrmion. 
Therefore, we have not only investigated the input spin texture but also the relative balance of $S_{\parallel}$ and $S_{\perp}$, whilst maintaining a fixed value for $S_{N}=$ 0.2~$\mu\mathrm{V}\cdot \mathrm{K}^{-1}$~\cite{Puttock2022}. 
Fig.~\ref{fig:main_two}(a) and (c) show analytically defined N\'{e}el and Bloch type skyrmions respectively, used as inputs to the thermoelectric calculation following the form~\cite{Romming2015, Barton2023}. 
Fig.~\ref{fig:main_two} (b)(i--iii) and (d)(i--iii) show the total thermoelectric response for the N\'{e}el and Bloch skyrmions respectively, where in each case (i), (ii) and (iii) correspond to $S_{\perp} =$ -0.50, -0.75, and -1.0~$\mu\mathrm{V}\cdot \mathrm{K}^{-1}$, while for every case $S_{\parallel}$ was held fixed at -1.0~$\mu\mathrm{V}\cdot \mathrm{K}^{-1}$. 
These values are selected to approximate the resulting thermoelectric signature, focusing on the spatial response as opposed to absolute values. 
We demonstrate that, for lower values of $S_{\perp}$ when the difference between $S_{\perp}$ and $S_{\parallel}$, the total thermoelectric response $V_{\mathrm{Total}}$ is significantly different between the two spin textures (Fig.~\ref{fig:main_two} (b)(i) and (d)(i) for instance). 
For the parameters simulated, the total signal is well within the sensitivity limits of the SThEM technique~($\approx$15~nV). 
We show that, as $S_{\perp}$ is increased, the spatial variation of $V_{\mathrm{Total}}$ for both the N\'{e}el and Bloch type skyrmions gradually approach equivalence, together with a gradual increase in the skyrmion signal amplitude, such that when the ratio $S_{\perp}/S_{\parallel} = 1$ (Fig.~\ref{fig:main_two} (b)(iii) and (d)(iii)), there is no difference between the signals of the two spin textures. 
We also highlight the fact that our model additionally captures the edge response observed in the experimental data. 
To expound on these observations, we separate the individual thermoelectric components of $V_{\mathrm{Total}}$ from Fig.~\ref{fig:main_two} (b)(i) and (d)(i). 
The relevant expressions for the ANE, PNE, and AMTP (Eqns.~\ref{Eqn:2}, ~\ref{Eqn:3} and ~\ref{Eqn:4}, respectively) are given by:     
\begin{widetext}
\begin{subnumcases}{V_y = }
        \int_{x}^{} \frac{dx}{w(x)}\int_{y}\left [ S_{N}\frac{m_{z}}{\left | m\right |} \right ]\nabla T_{x}(x_{0},y_{0}) dy, \label{Eqn:2}\\
				\int_{x}^{} \frac{dx}{w(x)}\int_{y}\left [ (S_{\parallel}-S_{\perp})\frac{m_{x}m_{y}}{\left | m\right |^{2}} \right ]\nabla T_{x}(x_{0},y_{0}) dy, \label{Eqn:3} \\
				\int_{x}^{} \frac{dx}{w(x)}\int_{y}\left [ S_{\perp} \left ( 1 - \frac{m_{y}^{2}}{\left | m\right | ^{2}} \right ) + S_{\parallel}\frac{m_{y}^{2}}{\left | m \right | ^2} \right ] \nabla T_{y}(x_{0},y_{0})dy \label{Eqn:4},
\end{subnumcases}
\end{widetext}
where $w(x)$ is the width of the device and $x_{0}$ and $y_{0}$ are $x$ and $y$ positions of the thermal probe as it is scanned during the thermoelectric calculation.\\
These expressions allow us to understand several features in the experimental and calculated data. 
The edge features arise purely due to the ANE (Eqn.~\ref{Eqn:2}) and is a consequence of the broken symmetry of the thermal gradient at the wire edge where the integrated effect of $\nabla T_{x}(x_{0},y_{0})$ is purely positive or negative. 
This leads to the positive or negative signal at the device edges where $\nabla T_{x}(x_{0},y_{0})$ is opposite in sign.
At the device centre there is an equal contribution such that the integrated response of $\nabla T_{x}(x_{0},y_{0})$ is nulled and no signal results from saturated moments, as shown in Fig.~\ref{fig:main_one}(d) and Fig.~\ref{fig:main_two} (b) and (d). 
We note that the gradient in the measured thermoelectric response across the device can be used to extract an estimate for the value for $S_{N}$.\\
Furthermore, we also observe the emergence of the ANE from the skyrmion itself, which is derived from $\nabla T_{x}(x_{0},y_{0})$ and the z-component of magnetisation, which is reversed with respect to the saturated region. 
As such, the thermoelectric response from the skyrmion is also reversed with respect to the edge response.
This is best observed in the data presented in Fig.~\ref{fig:main_two} (b)(iii) and (d)(iii), where $S_{\perp}$ and $S_{\parallel}$ are equal and therefore only the ANE response emerges.\\
\begin{figure*}[t!]
	\includegraphics[width=0.9\textwidth]{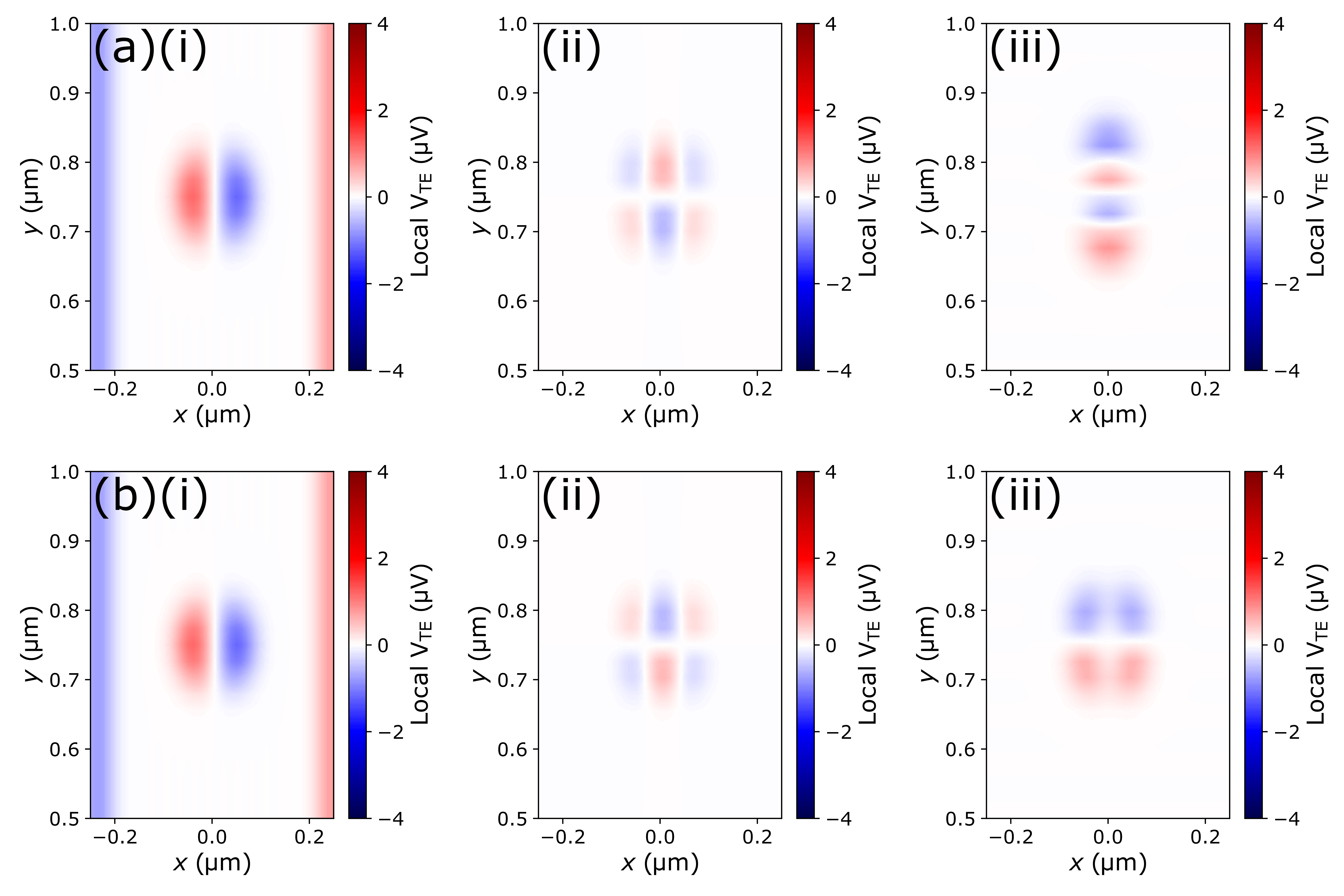}
	\caption{\label{fig:main_three} Individual components of the thermoelectric response from (a) N\'{e}el and (b) Bloch skyrmion spin textures respectively. 
	In each case, (i--iii) show the individual responses from the ANE, PNE and AMTP receptively. 
	$S_{\perp}$  was -0.5~$\mu\mathrm{V}\cdot \mathrm{K}^{-1}$ and $S_{\parallel}$ was -1.0~$\mu\mathrm{V}\cdot \mathrm{K}^{-1}$ for all calculations.}
\end{figure*}
To understand the nuances that arise due to the spins within the skyrmion's domain wall, we plot the individual components of the thermoelectric response from Eqn.~2 in Fig.~\ref{fig:main_three}. 
We show the individual thermoelectric response for the N\'{e}el and Bloch type skyrmion in Fig.~\ref{fig:main_three}(a) and (b), respectively, where in each case (i), (ii) and (iii) are the ANE, PNE, and AMTP, respectively. 
The ANE is identical for both N\'{e}el and Bloch as expected, due to the same z-component for each skyrmion type. 
When the in-plane spin components are considered, we see that the PNE and AMTP differ for the two skyrmion types. 
The PNE results in an antisymmetric multi-lobe pattern that is inverted for the two skyrmion types. 
This can be understood from the cyclic nature of the $x$ and $y$ components of the N\'{e}el and Bloch spin textures, which contribute to the $m_{x}m_{y}$ term in Eqn.~\ref{Eqn:3} and lead to the inverse relationship between the two. 
This cyclic response is broken for the AMTP as it solely depends on $m_{y}$; this leads to the differing responses observed in Fig.~\ref{fig:main_three} (a)(iii) and (b)(iii), respectively. 
From the individual responses, we can see that their sum leads to the results presented in Fig.~\ref{fig:main_two}, where the PNE and AMTP act to modulate the response of the ANE for the two skyrmion types, creating two distinct sets of features in $V_{\mathrm{Total}}$.\\
\begin{figure}[!ht]
\includegraphics[width=0.42\textwidth]{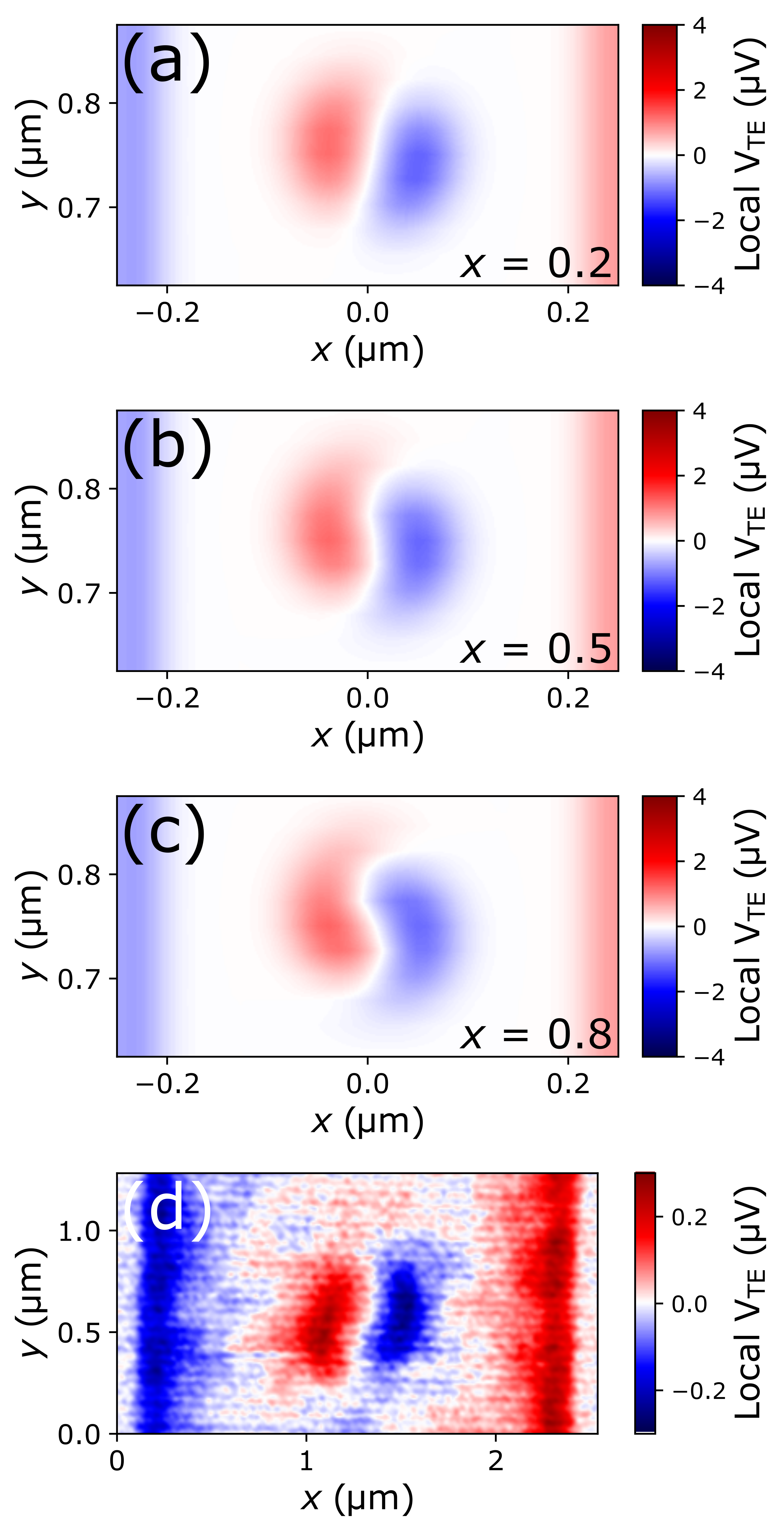}
\caption{Combined Bloch and N\'{e}el thermoelectric response, showing the normalised sum for a varying ratio of Bloch to N\'{e}el magnetisation configuration $V_{N\acute{e}el}x$ + $V_{Bloch}(1-x)$ for: (a) $x=$ 0.2, (b) $x=$ 0.5 and (c) $x=$ 0.8. \highlight{Panel (d) shows a zoom in of the experimental data from Figure~\ref{fig:main_one}(d) for comparison.}}
\label{fig:main_four}
\end{figure}
Qualitative comparison between these calculated thermoelectric responses and our experimental observations in Fig.~\ref{fig:main_one}(d) shows that there is a significant contribution from the ANE. 
Moreover, we observe a clear additional thermoelectric response from the skyrmion cross-sectional profile of the experimental data, see Fig.~\ref{fig:main_one}(e)(i). 
This arises due to the additional contribution from the PNE and AMTP, hence contains a thermoelectric response from the wall type. 
The \emph{Yin–Yangesque} nature of the skyrmion and the lower left~(red) lobe in Fig.~\ref{fig:main_one}(d) suggests a N\'{e}el--like texture where $S_{\parallel} - S_{\perp}$ is $\approx$0.25~$\mu \mathrm{V} \cdot \mathrm{K}^{-1}$. 
However, as the measured skyrmion is not a perfect circle~(magnetic inhomogeneities), and potential deviation from the idealised spin texture~\cite{Legrand2018} due to increased magnetostatic contribution in magnetic multilayers, it is difficult to conclude with absolute certainty.
\edit{Experimental measurements of the Seebeck coefficients are incredibly limited. However, recent studies on similar CoFeB/NM bilayer structures (where NM is a nonmagnetic metal known to produce DMI such as Pt, Ta, Pd etc.) have reported measurements of the spin Seebeck coefficient of between 0.275--1.13~$\mathrm{\mu V \cdot K^{-1}}$, depending on the specific choice of NM~\cite{Lee2015, GAMINO2021167778}. This, together with the calculated anisotropy of the Seebeck coefficient of between 0.1--0.3~$\mathrm{\mu V \cdot K^{-1}}$ (depending on temperature and multilayer repetitions), suggests that the parameters used in the calculations are a reasonable assumption~\cite{PhysRevB.88.104425}.}
Extending this idea further, we have computed the normalised sum of fractional thermoelectric contributions from the Bloch and N\'{e}el respectively. 
Here, we calculate the sum $V_{N\acute{e}el}x$ + $V_{Bloch}(1-x)$ for a range of $x$ values. 
Fig.~\ref{fig:main_four} shows the calculated normalised sum for three values of $x$: 0.2, 0.5 and 0.8, (a)--(c) respectively and where $S_{\perp}=$ 0.5 $\mu \mathrm{V} \cdot \mathrm{K}^{-1}$.  
Here, we demonstrate how the overall response can be tuned from N\'{e}el like to Bloch like through an average of the two spin textures when $x=$ 0.5. 
This simple approach allows us to ascertain how the total $V_{TE}$ could vary for non--ideal spin textures and demonstrates how a mixed state could approximate the experimental data. 
We note that our stack comprises of three trilayers and therefore, is below the typical repetition number~\cite{Legrand2018} for magnetostatics to lead to hybrid skyrmion textures. 
Therefore, we propose that an alternative explanation is that $S_{\parallel} - S_{\perp}$ is small but non--negligible. 
In this scenario, the difference is such that the peak contribution from the skyrmion domain wall is detectable within the sensitivity limits of the technique, as demonstrated in line profile through the experimental data Fig.~\ref{fig:main_one}(e)(i). 
However, the spatial extent and the tails present in the N\'{e}el skyrmion, Fig.~\ref{fig:main_two} (b)(ii), are such that they are below the detection limit of the measurement, reducing the distinguishability in this case. \highlight{This represents an interesting avenue to explore in future work. Scanning thermoelectric microscopy could be performed on thin lamellae of B20 crystals or 2D ferromagnets, that have been shown to be room-temperature or near-room-temperature Bloch skyrmion hosts~\cite{Twitchett-Harrison2022, Lv2024}. Additionally, the results presented in Fig.~\ref{fig:main_four} could be used to interpret skyrmions in large-repetition multilayers, where it has been shown that skyrmions possess mixed Bloch-Neel characteristics~\cite{Legrand2018}.}\\         
Nevertheless, our modelling shows that SThEM is a powerful technique for fingerprinting magnetic spin textures and has the potential to reveal the N\'{e}el or Bloch nature of individual skyrmions using a simple laboratory setup. 
We suggest it would be best applied to ideal spin textures such at those found in two-dimensional ferromagnets~(pure spin textures), and for materials systems where $S_{\perp}$ and $S_{\parallel}$ are significantly different, increasing the signal--to--noise ratio. 
\highlight{Inspection of Eqns. (2) and (3) show that because the $m_x m_y$ product is the same in either case, it is not possible to determine the helicity from SThM alone. Similarly, it was shown that it was not possible to measure the chirality of Néel walls with this technique for the same reason~\cite{Puttock2022}. However, SThM analysis can easily be combined with MFM, often in the same microscope set-up. It has been shown that, with MFM, the helicity of skyrmions can be determined~\cite{Barton2023}, making the combination of these techniques ideal for fast characterisation of spin textures.}
Finally, although considering the TE response due to inplane thermal gradients captures the experimentally measured data well, it is conceivable that the vertical gradient could yield additional information. 
Indeed, our Seebeck tensor already reveals this relationship for the effects and experimental geometry considered here. 
Therefore, we anticipate future work would involve accounting for any potential contribution due to $\nabla T_{z}$ and subsequent modification of the Seebeck tensor to account for additional effects such as the spin Seebeck effect~(SSE)--particularly important for insulating magnetic samples~\cite{Uchida2008, Sola2020}.\\
To conclude, we have used SThEM to measure the local thermoelectric response from a single skyrmion when interrogated with a highly localised thermal gradient. 
The resultant thermoelectric response is the sum of the individual components from all thermally driven spin dependent transport phenomena. 
These are included in the Seebeck tensor which is used to derive various expressions for the individual thermoelectric voltages. 
Therefore, the measured voltage is directly linked to the underlying spin structure of the skymrion. 
Through additional thermoelectric modelling, we show it possible to fingerprint skyrmion types, Bloch and N\'{e}el, due to the unique thermoelectric responses resulting from their differing underlying spin textures. 
The extent to which their spin textures can be distinguished is explored by varying the relative size of the in-plane Seebeck coefficients, $S_{\perp}$ and $S_{\parallel}$, which is material dependent. 
Our study demonstrates how a simple atomic force microscope can be harnessed to explore the underlying magnetisation of topological spin textures. 
We further suggest, based on our best-case measurement noise floor of $\approx$~15 nV, that this methodology could be employed to also explore thermally driven topological signatures in the measured data. 
The combination of high sensitivity and position-control of the stimulus in such proximity to the spin texture could create a fruitful playground for highly localised investigations of topological phenomenon. 
Our work will open a route to nanoscale optimisation of skyrmion-based spin caloritronic materials and devices, accelerating existing technology readiness, and additionally driving progress in the next generation of devices where thermally driven topological phenomenon will play a central role.\\
\section*{Supplementary Material}
\highlight{See supplementary material for additional scanning thermoelectric microscopy images of zero field skyrmions in devices fabricated from both identical and similar multilayers and example calculated thermal gradient used in the modelling of the thermoelectric response.}
\begin{acknowledgments}
This work was supported in part by the European Metrology Research Programme (EMRP) and EMRP participating countries under the European Metrology Programme for Innovation and Research (EMPIR) Project No. 17FUN08 TOPS: Metrology for topological spin structures. 
The project also received financial support from the UK government department for Science, Innovation and Technology through NMS funding (Metrology of complex systems for low energy computation). 
Financial support was also received through the PID2022-137567NB-C21/AEI/10.13039/501100011033 grant from the Spanish Ministry of Science and Innovation. 
Further support was also received through European Union Horizon 2020 funding scheme--(H2020 grant MAGicSky No. FET-Open-665095.103)
\end{acknowledgments}              
 
\section*{\label{sec:AuthorDeclarations}Author Declarations}

\subsection*{\label{sec:CoI}Conflict of Interest}
The authors declare no conflicts of interest.

\subsection*{\label{sec:AuthorCont}Author Contributions}
\textbf{Christopher E. A. Barker}: Conceptualization~(equal); Formal analysis~(equal); Methodology~(equal); Writing -- original draft (lead); Writing -- review \& editing~(equal); \textbf{Elias Saugar}: Conceptualization~(equal); Formal analysis~(equal); Methodology~(equal); Writing -- review \& editing~(equal); \textbf{Robert Puttock}: Conceptualization~(supporting); Formal analysis~(supporting); Methodology~(supporting); Writing -- review \& editing~(equal); \textbf{Katharina Zeissler}: Conceptualization~(supporting); Formal analysis~(supporting); Methodology~(supporting); Writing -- review \& editing~(equal); \textbf{Petr Klapetek}: Conceptualization~(supporting); Formal analysis~(supporting); Methodology~(supporting); Writing -- review \& editing~(equal); \textbf{Olga Kazakova}: Conceptualization~(supporting); Formal analysis~(supporting); Methodology~(supporting); Writing -- review \& editing~(equal); \textbf{Christopher H. Marrows}: Conceptualization~(supporting); Formal analysis~(supporting); Methodology~(supporting); Supervision~(lead); Writing -- review \& editing~(equal); \textbf{Oksana Chubykalo-Fesenko}: Conceptualization~(lead); Formal analysis~(lead); Funding acquisition~(lead); Methodology~(equal); Supervision~(lead); Writing -- review \& editing~(equal); \textbf{Craig Barton}: Project administration~(lead); Conceptualization~(lead); Formal analysis~(lead); Funding acquisition~(lead); Methodology~(lead); Supervision~(lead); Writing -- original draft (lead); Writing -- review \& editing~(equal).

\subsection*{\label{sec:Data}Data Availability}
The data that support the findings of this study are available within the following online repository \ldots

      
\bibliography{MainText-refs}

\end{document}